\documentclass[11pt, conference]{IEEEtran}
\IEEEoverridecommandlockouts
\usepackage{cite}
\usepackage{graphicx}
\usepackage{soul}
\usepackage{color}
\usepackage{xcolor}
\graphicspath{ {./images/} }
\usepackage{amsmath,amssymb,amsfonts}
\usepackage{algorithmic}
\usepackage{graphicx}
\usepackage{textcomp}
\usepackage{xcolor}
\usepackage{blindtext}
\usepackage{hyperref}
\begin{document}

\title{CrowdWeb: A Visualization Tool for Mobility Patterns in Smart Cities}

\author{\IEEEauthorblockN{Yisheng Alison Zheng,  Abdallah Lakhdari, Amani Abusafia, Shing Tai Tony Lui and Athman Bouguettaya}
\IEEEauthorblockA{School of Computer Science\\
 The University of Sydney, Australia\\	
\{yzhe5242, abdallah.lakhdari,  amani.abusafia, slui2950, athman.bouguettaya\}@sydney.edu.au}
}
%Shing Tai Tony Lui
%slui2950
\maketitle

\begin{abstract}

Human mobility patterns refer to the regularities and trends in the way people move, travel, or navigate through different geographical locations over time. Detecting human mobility patterns is essential for a variety of applications, including smart cities, transportation management, and disaster response.  The accuracy of current mobility prediction models is less than 25\%. The low accuracy is mainly due to the fluid nature of human movement. Typically, humans do not adhere to rigid patterns in their daily activities, making it difficult to identify hidden regularities in their data. To address this issue,  we proposed a web platform to visualize human mobility patterns by abstracting the locations into a set of places to detect more realistic patterns. However, the platform was initially designed to detect individual mobility patterns, making it unsuitable for representing the crowd in a smart city scale.  Therefore, we extend the platform to visualize the mobility of multiple users from a city-scale perspective. Our platform allows users to visualize a graph of visited places based on their historical records using a modified PrefixSpan approach. Additionally, the platform synchronizes, aggregates, and displays crowd mobility patterns across various time intervals within a smart city. We showcase our platform using a real dataset.%\looseness=-1
%Our platform enables users to visualize a graph of the places they visited based on their history records using a modified PrefixSpan approach. Moreover, the platform aggregates and presents the crowd mobility patterns over different time intervals in a smart city. We demonstrate our platform using a real dataset.

\end{abstract}


%\textit{Human Mobility patterns} are the sequences of frequently visited places by an individual \cite{gonzalez2008understanding}. Detecting the mobility patterns of individuals is essential in several fields such as pandemic prevention \cite{eubank2004modelling,wang2019urban}, urban planning \cite{xia2018exploring} crowd management \cite{higashino2018re,wang2017online,zhou2020understanding}, and location-based services \cite{karamshuk2011human}.  Several studies have shown that human mobility is highly predictable, and this can be attributed to the regularity of our daily routines \cite{gonzalez2008understanding,yang2014modeling}. The process of acquiring human mobility patterns involves analyzing the spatio-temporal attributes and potential regularities in individual and population movement trajectories \cite{wang2019urban,solmaz2019toward}. Several  models have been proposed to represent and predict human mobility patterns \cite{haifeng2021human}. The availability of location-based data through social networks has provided a unique opportunity to explore these patterns in depth from a quantitative and microscopic perspective \cite{wang2019urban}.

\textit{Human Mobility Patterns} are the series of places frequently visited by an individual \cite{gonzalez2008understanding}. Detecting these mobility patterns is crucial in various domains such as pandemic prevention \cite{wang2019urban}, urban planning \cite{xia2018exploring}, crowd management \cite{higashino2018re,Amani2022QoE}, and location-based services \cite{yang2023monitoring,lakhdari2016link}. Several studies have demonstrated that human mobility is highly predictable due to the regularity of daily routines \cite{gonzalez2008understanding,zhou2018understanding}. The acquisition of human mobility patterns involves examining spatio-temporal attributes and uncovering potential regularities in individual and population movement trajectories \cite{wang2019urban,solmaz2019toward}. Several models have been proposed to represent and predict human mobility patterns \cite{haifeng2021human}. The availability of location-based data through social networks offers a unique opportunity to comprehensively investigate these patterns from both quantitative and detailed perspectives \cite{wang2019urban}.\looseness=-1

The detection of individual mobility patterns necessitates analyzing the historical data of the places visited by individuals. Deep learning approaches have been suggested for forecasting a user's next point of interest \cite{luca2021survey, haifeng2021human}; however, these methods have limited accuracy, ranging from 8\% to 25\%. Detecting an exact mobility pattern for a user is challenging because of the inherent flexibility of human movement \cite{gonzalez2008understanding,wang2019urban}. For example, a user who regularly eats Thai food for lunch between 12:00 and 13:00 may visit a different Thai eatery each day, e.g., Thai Express on the first day, Seasoning Thai on the second day, and Thai Pothong on the third day. Despite the user's consistent dining habits, it is difficult to recognize this pattern due to the varying locations of the restaurants. Consequently, we proposed a platform for visualizing the mobility patterns of labeled locations to more accurately define users' mobility patterns \cite{zheng2022imap}.% For example, labeling the three Thai eateries as ``Thai restaurant" would enable the detection of a more accurate pattern.   
The platform displays a set of \textit{frequent mobility patterns}  by using a modified PrefixSpan algorithm \cite{pei2004mining}. However, the platform was designed to detect individual mobility patterns, making it unsuitable for representing a group of users (i.e., a crowd) on a smart-city scale.\looseness=-1

\begin{figure}[!t]
    \centering
    \setlength{\abovecaptionskip}{0pt}
    \setlength{\belowcaptionskip}{-20pt}
    \includegraphics[width=0.9\linewidth]{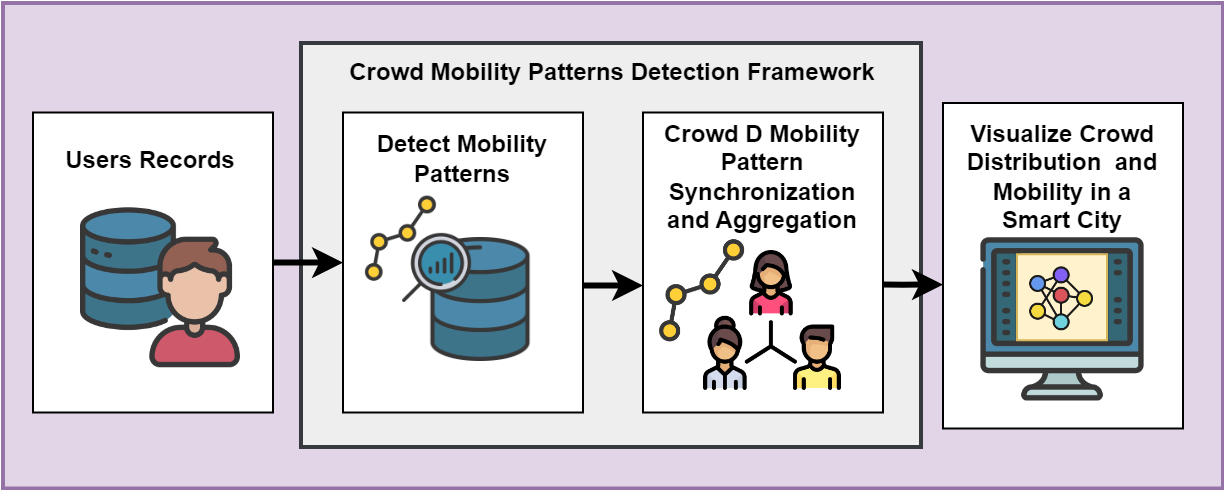}
    \caption{System overview}
    \label{fig:system}
\end{figure}

 Detecting crowd mobility in a smart city is crucial in applications such as crowd management \cite{higashino2018re,abusafia2022maximizing,zhou2020understanding}, IoT services \cite{yang2022towards,yao2022wireless} and pandemic control \cite{wang2019urban}. Detecting crowd mobility is challenging because people have different spatio-temporal patterns \cite{zheng2022imap,abusafia2022service}.  In this paper, we extend the aforementioned platform to compute and visualize the patterns of a crowd in a smart city over various time periods (See Fig.\ref{fig:system}).  The platform utilizes users' records after labeling the locations and their computed \textit{mobility patterns} to compute the crowd mobility patterns and distribution  (See Fig.\ref{fig:user1}).  The mobility patterns of each user are detected using a modified PrefixSpan algorithm \cite{pei2004mining}.  Our platform synchronizes the mobility patterns of crowds by identifying a map of their visited places (see Fig.\ref{fig:user1}. Moreover, we align the crowd's patterns using different time windows.  Our platform facilitates the analysis and comprehension of a crowd's movement within a city.\looseness=-1

% We present, \textit{IMAP}, an Individual huMAn mobility Patterns visualizing platform. Our platform enables users to \textit{ visualize a graph} of the places they visited based on their history records. In addition, our platform displays the most \textit{frequent mobility patterns} computed using a modified PrefixSpan approach.

%%\vspace{-14pt}
\section{Crowd Mobility Patterns Detection Framework}
%%\vspace{-5pt}

\begin{figure}[!t]
    \centering
    \setlength{\abovecaptionskip}{0pt}
    \setlength{\belowcaptionskip}{-20pt}
    \includegraphics[width=\linewidth]{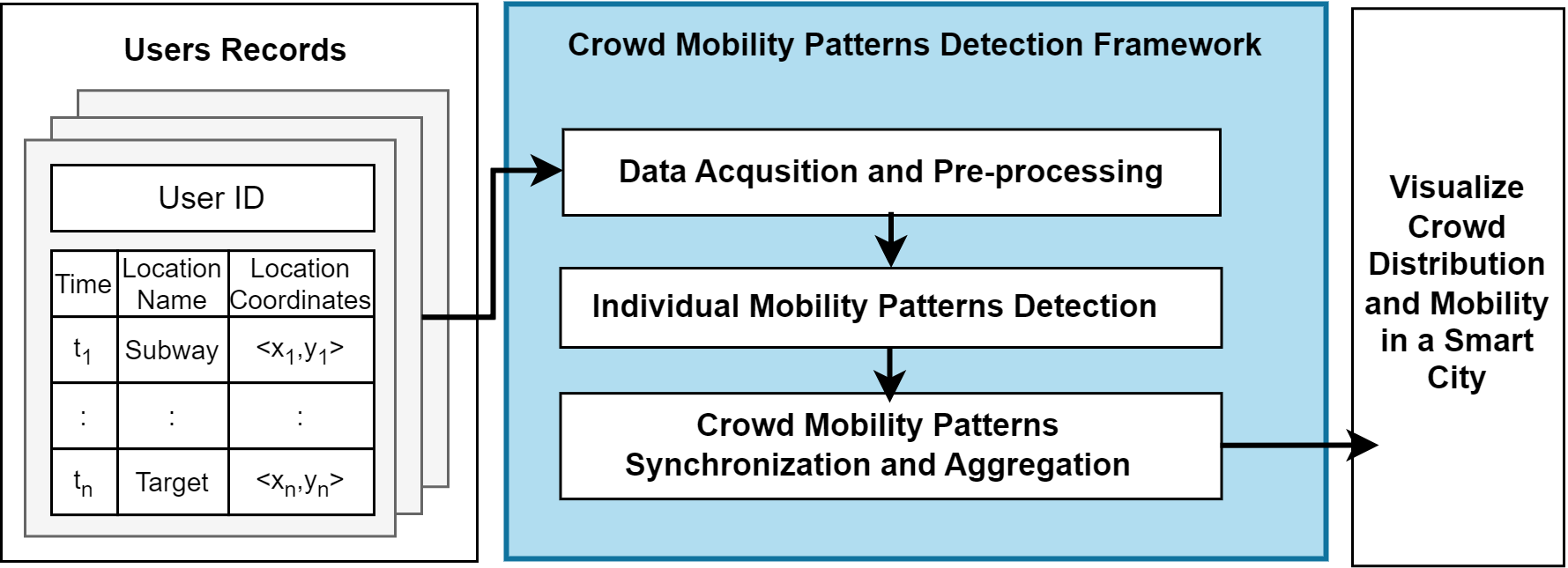}
    \caption{Crowd Mobility Patterns Detection Framework}
    \label{fig:framework}
\end{figure}

The crowd mobility patterns detection framework aims to synchronize, aggregate, and display crowd distributions and mobility patterns across various time intervals within a smart city. The framework comprises three phases: data acquisition and pre-processing, individual mobility patterns detection, and crowd distribution and mobility patterns synchronization and aggregation (See Fig. \ref{fig:framework}). In what follows, we discuss each phase in details:%\looseness=-1
%Our platform enables users to visualize a single user's mobility graph and patterns. Users have the option to choose the default dataset or upload their check-in history from a social network platform such as Foursquare. Furthermore, the platform allows users to visualize the mobility of a crowd in a smart city at specific times using the same social network dataset. In this demonstration, we employed the Foursquare dataset as our default dataset \cite{yang2014modeling}. The Foursquare dataset is a \textit{geo-tagged social media} (GTSM) dataset where users check in at venues they visit. We utilized the New York dataset, which contains 227,428 check-in records. The platform consists of three primary steps (refer to Fig.\ref{fig:system}): %%looseness=-1"

\subsubsection{Data Acquisition and Pre-processing}

During this phase, users' information and daily visited places data are acquired from a geo-location dataset. In this demonstration, we used a public Foursquare dataset as our default dataset \cite{zheng2022imap}. The Foursquare dataset is a \textit{geo-tagged social media} (GTSM) dataset, where users check in at the venues they visit. We used the New York dataset, which comprises 227,428 check-in records. The dataset was collected over an 11-month period (April 2012 to February 2013). As the GTSM dataset is collected by allowing users to check in voluntarily, it is possible that users are not checking in regularly. This leads to a sparse dataset. To confirm this, the average number of records for each user is examined. The average is approximately 210, and the median is 153. As there are approximately 330 days in the data collection period, there would be less than one record per day. Hence, the dataset is sparse. In addition, there are 1083 users in the dataset. Therefore, to address the data sparsity, we aimed to extract data from months with rich check-in records. After investigating, we found that the best month is the period of time from April to June. Therefore, for the experiment, we employed data from these three months. Moreover, we discovered that users were not recording their movement patterns on a daily basis. However, in order to extract a descriptive human mobility pattern for the users, we would need to ensure the user records are rich. Hence, we selected users with less than 2 hours check-in records for more than 50 days within the 3-month period.\looseness=-1

\looseness=-1

\subsubsection{Individual Mobility Patterns Detection} 
As presented in \cite{zheng2022imap}, a modified PrefixSpan algorithm is used to detect the mobility patterns of each user \cite{pei2004mining}. %
%the user records of visited places are relabeled by the name of these places. For instance, all Thai restaurants will be labeled as ``Thai restaurant". The labeling rules can be determined by domain experts based on the purpose of the pattern mining. After updating the users' records, a modified PrefixSpan algorithm is used to detect the mobility patterns of each user \cite{pei2004mining}.
    
\subsubsection{Crowd Mobility Patterns Synchronization and Aggregation}
\begin{figure}[!t]
    \centering
    \setlength{\abovecaptionskip}{0pt}
    \setlength{\belowcaptionskip}{-15pt}
    \includegraphics[width=0.8\linewidth]{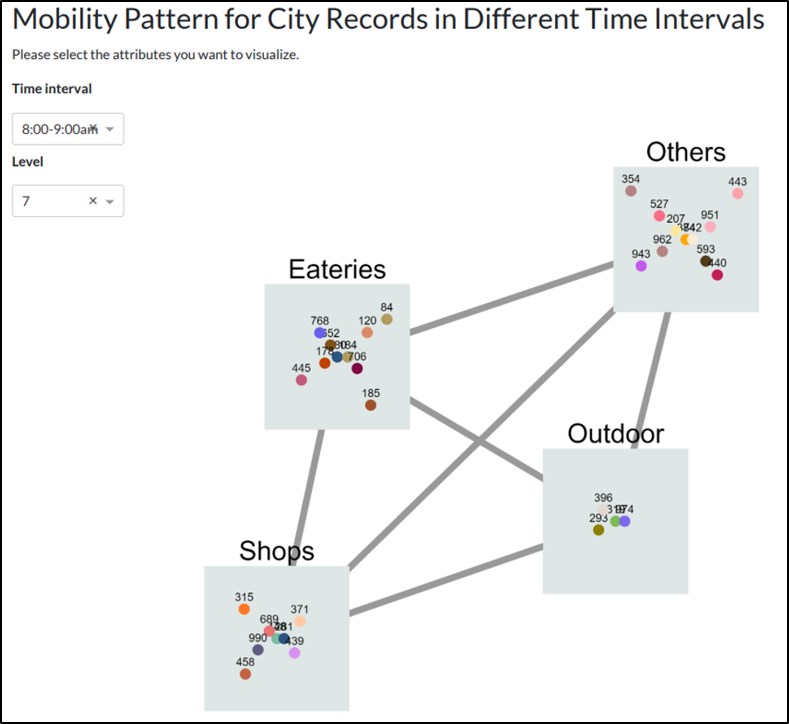}
    \caption{The crowd in a smart city from 9-10 am}
    \label{fig:user1}
\end{figure}

\begin{figure}[!t]
    \centering
    \setlength{\abovecaptionskip}{0pt}
    \setlength{\belowcaptionskip}{-20pt}
    \includegraphics[width=0.8\linewidth]{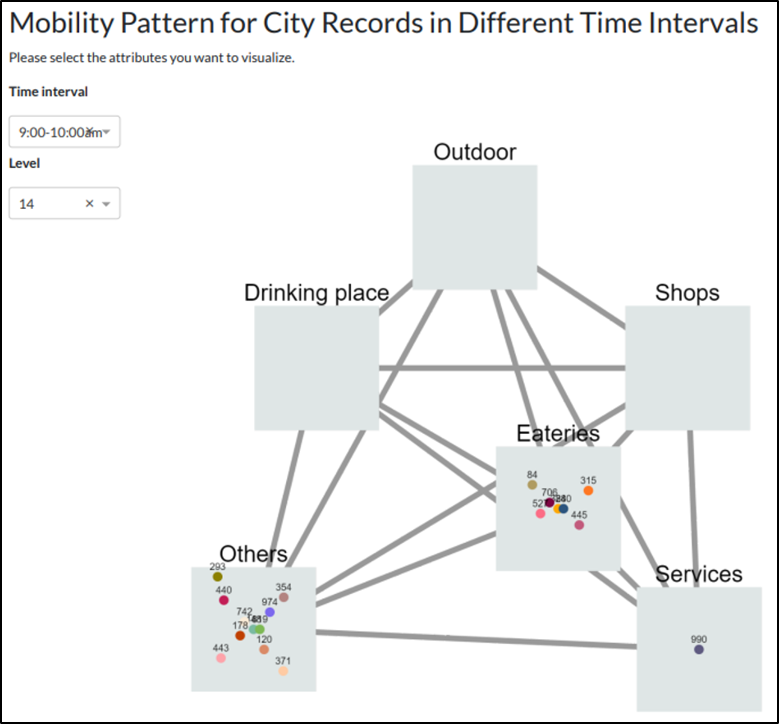}
    \caption{The crowd in a smart city from 9-10 am}
    \label{fig:user2}
\end{figure}

In this step, we synchronize and aggregate all the users' mobility patterns based on time. Users who frequently visit a specific location at a particular time are categorized together as a group (See Fig.\ref{fig:user1}). For example, any user with a pattern of visiting a certain microcell (e.g., shops) at a certain selected time (e.g., 8:00 am) will appear in the smart city at the selected time  (See Fig.\ref{fig:user1}). Moreover, if we change the time, the crowd locations may change to other microcells, depending on their patterns (See Fig.\ref{fig:user2}). Our platform uses the aggregated patterns and distributions to visualize the crowd movement in a smart city.\looseness=-1

%%looseness=-1

% %%looseness=-1 .    
% \textbf{Visualize the mobility graph and patterns:} In this step, we visualize the crowd mobility patterns in a smart city at a selected time.  

\section{Demo Setup}
% Our demo presents an interactive web application for a crowd in a smart city. Additionally, we will display a recorded video of the entire process of using the platform to present and interact with a group of default users' patterns in real-time. Visitors to our booth can choose from a list of available users to visualize their network and their mobility patterns. They can also choose the city visualization and visualize the crowd movement over different time windows. If any of the audience is willing to share their history, we may upload it to the platform and visualize their patterns.  

Our interactive web application demonstrates crowd mobility in a smart city. We will also present a recorded video that presents the entire process of using the platform to display and interact with the default users' patterns in real-time. The video can be found at this link: \href{https://tinyurl.com/crowdweb}. Booth visitors can select from a list of available users to visualize their networks and mobility patterns. Additionally, they can choose the city visualization and observe crowd movements across various time frames. If any audience member is willing to share their check-in history, we can upload it to the platform and visualize their patterns.\looseness=-1

\section{Mobility Patterns Exploration}
% \vspace{-5pt}
% To explore the detected mobility patterns using the Foursquare dataset, we evaluated the effect of the Minimum Support Threshold  $min\_support$ on the patterns detected by the modified PrefixSpan \cite{zheng2022imap}.  First, we checked the effect of changing  $min\_support$  on the number of sequences extracted per user. Second, we check the effect of changing  $min\_support$  on the average length of the sequences extracted per user.\looseness=-1

To investigate the detected mobility patterns using the Foursquare dataset, we evaluated the effect of the Minimum Support Threshold $min\_support$ on the patterns detected by the modified PrefixSpan \cite{zheng2022imap}. Firstly, we examined the effect of changing $min\_support$ on the number of sequences extracted per user. Secondly, we assessed the effect of changing $min\_support$ on the average length of the sequences extracted per user.\looseness=-1

\begin{figure}[!t]
    \centering
    \setlength{\abovecaptionskip}{0pt}
    \setlength{\belowcaptionskip}{0pt}
    \includegraphics[width=\linewidth]{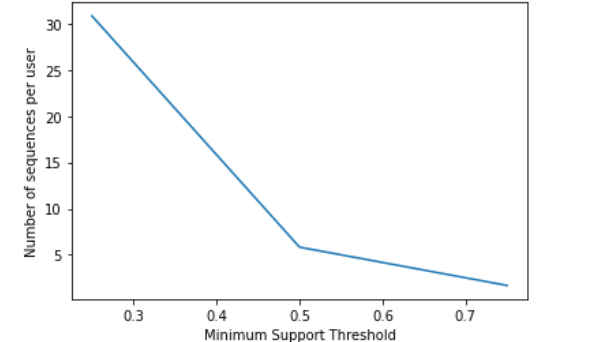}
    \caption{Average number of sequences per user vs. minimum support threshold }
    \label{fig:nth}
\end{figure}

\begin{figure}[!t]
    \centering
    \setlength{\abovecaptionskip}{0pt}
    \setlength{\belowcaptionskip}{-15pt}
    \includegraphics[width=0.8\linewidth]{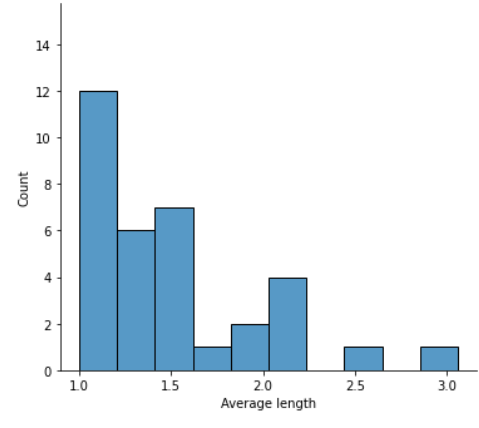}
    \caption{Distribution plot of average length with $min\_support$ = 0.5}
    \label{fig:dplot}
\end{figure}

The first experiment examined the effects of changing  $min\_support$  on the number of sequences extracted per user. Figure \ref{fig:nth} illustrates the correlation between the number of sequences per user and minimum support threshold. In general, the number of sequences per user decreased as the minimum support threshold increased. This trend occurs because a higher minimum support threshold value makes it more difficult for a pattern to be recognized as a sequential pattern. It is also notable that, as the minimum support threshold increases from 0.25 to 0.5, there is a significant decrease in the number of sequences per user. Conversely, when the minimum support threshold rises from 0.5 to 0.75, the decline in the number of sequences per user is less pronounced. To ascertain the previous evaluation, we present the distribution of the number of sequences with $min\_support$ = 0.5 in figure \ref{fig:dplot}.\looseness=-1

%The first experiment investigated the effects of altering $min\_support$ on the number of sequences extracted per user. Figure \ref{fig:nth} displays the relationship between the number of sequences per user and the minimum support threshold. Generally, the number of sequences per user declines as the minimum support threshold increases. This trend occurs because a higher minimum support threshold value makes it more challenging for a pattern to be identified as a sequential pattern. Interestingly, when the minimum support threshold increases from 0.25 to 0.5, there is a substantial decrease in the number of sequences per user. However, as the minimum support threshold rises from 0.5 to 0.75, the reduction in the number of sequences per user is comparatively smaller. Furthermore, Figure \ref{fig:dplot} illustrates the distribution of the number of sequences.

\begin{figure}[!t]
    \centering
    \setlength{\abovecaptionskip}{0pt}
    \setlength{\belowcaptionskip}{0pt}
    \includegraphics[width=\linewidth]{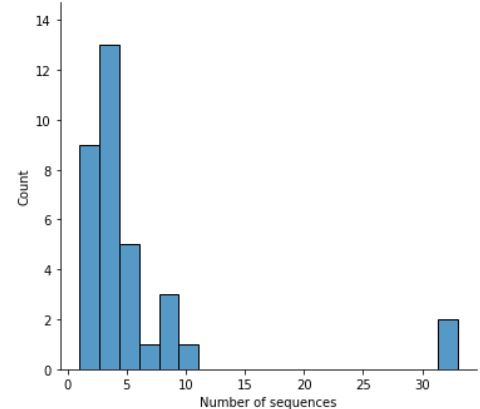}
    \caption{Distribution plot of the number of sequences with $min\_support$ = 0.5}
    \label{fig:dplot2}
\end{figure}

% Figure \ref{sgg} describes the relationship between the average length of sequences per user and the minimum support threshold. Generally, the average length of sequences per user decreases as the support threshold is increasing. Moroever,  When the minimum support threshold increases, the chance of a longer pattern to be certify as a sequential pattern will be much lower than that of a shorter pattern. For example, it is logical that a pattern 'Eatery' will be appearing more than that of the pattern 'Eatery, Shops' in a sequence database, hence there is a higher chance for the pattern 'Eatery' to be certify as a sequential pattern than that of 'Eatery, Shops'.

\begin{figure}[!t]
    \centering
    \setlength{\abovecaptionskip}{0pt}
    \setlength{\belowcaptionskip}{-15pt}
    \includegraphics[width=\linewidth]{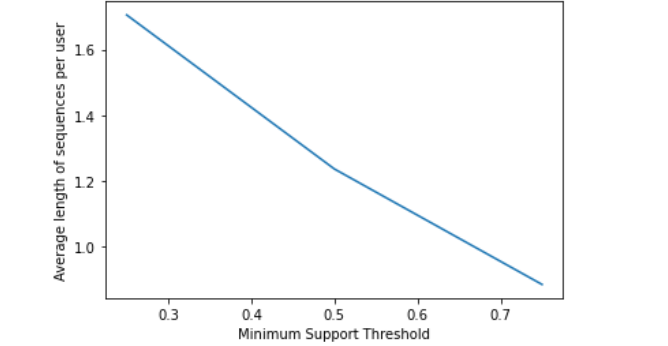}
    \caption{Average length of sequences per user vs. minimum support threshold}
    \label{fig:sgg}
\end{figure}

The second experiment examined the impact of modifying $min\_support$ on the average length of the sequences extracted per user. Figure \ref{fig:sgg} illustrates the relationship between the average length of sequences per user and the minimum support threshold. Generally, the average length of sequences per user decreases as the support threshold rises. Moreover, when the minimum support threshold increased, the likelihood of a longer pattern being recognized as a sequential pattern was considerably lower than that of a shorter pattern. For instance, it is reasonable to expect that the pattern 'Eatery' would appear more frequently than the pattern 'Eatery, Shops' in a sequence database, leading to a higher probability of 'Eatery' being certified as a sequential pattern compared to 'Eatery, Shops'. To ascertain the previous evaluation, we present the distribution of the average length across the levels with $min\_support$ = 0.5 in figure \ref{fig:dplot2}.\looseness=-1

 % In this demonstration, we employed the Foursquare dataset as our default dataset \cite{yang2014modeling}. The Foursquare dataset is a \textit{geo-tagged social media} (GTSM) dataset where users check in at venues they visit. We utilized the New York dataset, which contains 227,428 check-in records.

%%\vspace{-10pt}

\section{Conclusion}
This paper presents a web platform that visualizes human mobility patterns at both the individual and city-scale levels, making it more suitable for applications within a smart city scale.% By abstracting locations into a set of places and utilizing a modified PrefixSpan approach, the platform uncovers more realistic patterns of human movement. 
In addition to individual mobility patterns, the platform enables the visualization of multiple users' mobility patterns, offering a city-scale perspective. In the future, we plan to allow users to scale the time frames for the crowd movement and automate the crowd movement animation.\looseness=-1

\section*{Acknowledgment}
This research was partly made possible by LE220100078 and DP220101823 grants from the Australian Research Council. The statements made herein are solely the responsibility of the authors.

\bibliographystyle{IEEEtran}
\bibliography{main}

\end{document}